\documentstyle[11pt,epsf]{article}

\begin{document}
\title{How useful 
can knot and number theory be for loop
calculations?\thanks{Talk presented at the workshop {\em 
Loops and Legs in Gauge Theories}, Rheinsberg 1998,
Mainz Univ.~preprint {\bf MZ-TH/98-20}.}}
\author{Dirk Kreimer\\
Physics Dept., Mainz University, D-55099 Mainz\\
email: dirk.kreimer@uni-mainz.de}
\maketitle
\begin{abstract}
We summarize recent results connecting multiloop Feynman diagram
calculations to different parts of mathematics, with special attention
given
to the Hopf algebra structure of renormalization.
\end{abstract}
\section{Introduction}
In this contribution I want to report on some recent developments in the area
of multiloop Feynman diagram calculations. These developments touch widely
separated areas of mathematics, ranging from knot theory to number
theory 
\cite{habil} as well as from
combinatorical Hopf algebras \cite{hopf}
to the realm of noncommutative geometry \cite{alain,CM}
and the classification of operator algebras. 
At the same time, on the physics side, they touch
areas as separated as the Ising model \cite{DaI}
in three dimensions, the $\beta$-function in $\phi^4$
theory in four dimensions
\cite{pisa}, and  the recent identification of the unknown constant
in the $\rho$-parameter of the Standard Model \cite{DaM}.


In David Broadhurst's recent work you will find results concerning the identification
of transcendental numbers arising from such finite parts of Feynman diagrams
with the hyperbolic covolume of hyperbolic knots \cite{DaI,DaM,DaH}. Such results,
generalizing the by now familiar identification of knots, numbers and counterterms
\cite{habil,pisa,bk15,bgk,bdk,plb,book}, point towards a deep and yet to be clarified connection 
between QFT and
(hyperbolic) geometry.


The recent results of Andrei Davydychev and Bob Delbourgo \cite{ADRD} identifying
one-loop triangles and boxes with the geometry of the hyperbolic tetrahedron
point in the same direction. All these results give testimony to the still premature
and incomplete understanding of QFT which we still have more than half a century after its
discovery. On the other hand 
they give hope that we begin to see some parts
of the mathematics which underlies QFT.






Let me summarize the results achieved in recent years, and let me stress the most 
promising facts which deserve, in my opinion, future attention.





\section{Knots, Numbers, Diagrams}
Let us start with the consideration of a Feynman diagram which has a non-vanishing
superficial degree of divergence, but has no subdivergent graphs.
The coefficient of the divergence is then a well-defined number,
and independent of the chosen regularization or renormalization scheme.





\begin{figure}
\epsfysize=5cm\epsfbox{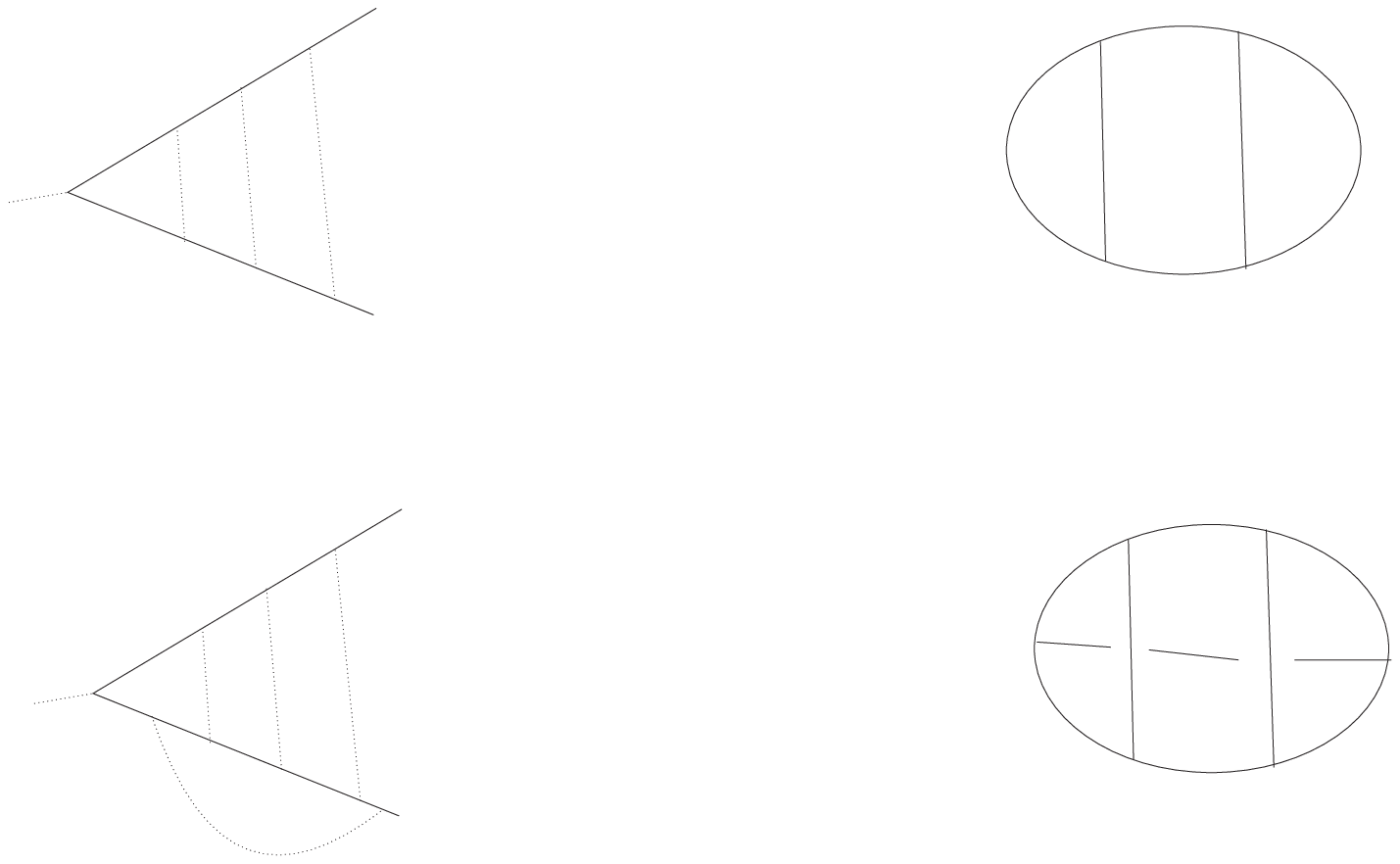}
\caption{The topology of a diagram. We discard external lines and the
  different nature of internal propagators. The resulting graph
solely encodes the topology of the diagram. This is sufficient
information
to determine the transcendental nature of the counterterm
of the Feynman graph.}
\label{f1}
\end{figure}

The study of such coefficients allows to observe some remarkable patterns:
we can abstract from the nature of the particles realizing a given topology, and consider
solely the topology of the graph as determined by internal propagators.
Whenever the topology of two different graphs matches,
even in different theories, we expect the same transcendental
number from these graphs. 
In Fig.(\ref{f1}) we see some graphs and their topology
wich completely determines the transcendentals in their counterterms.

\begin{figure}
\epsfysize=12cm\epsfbox{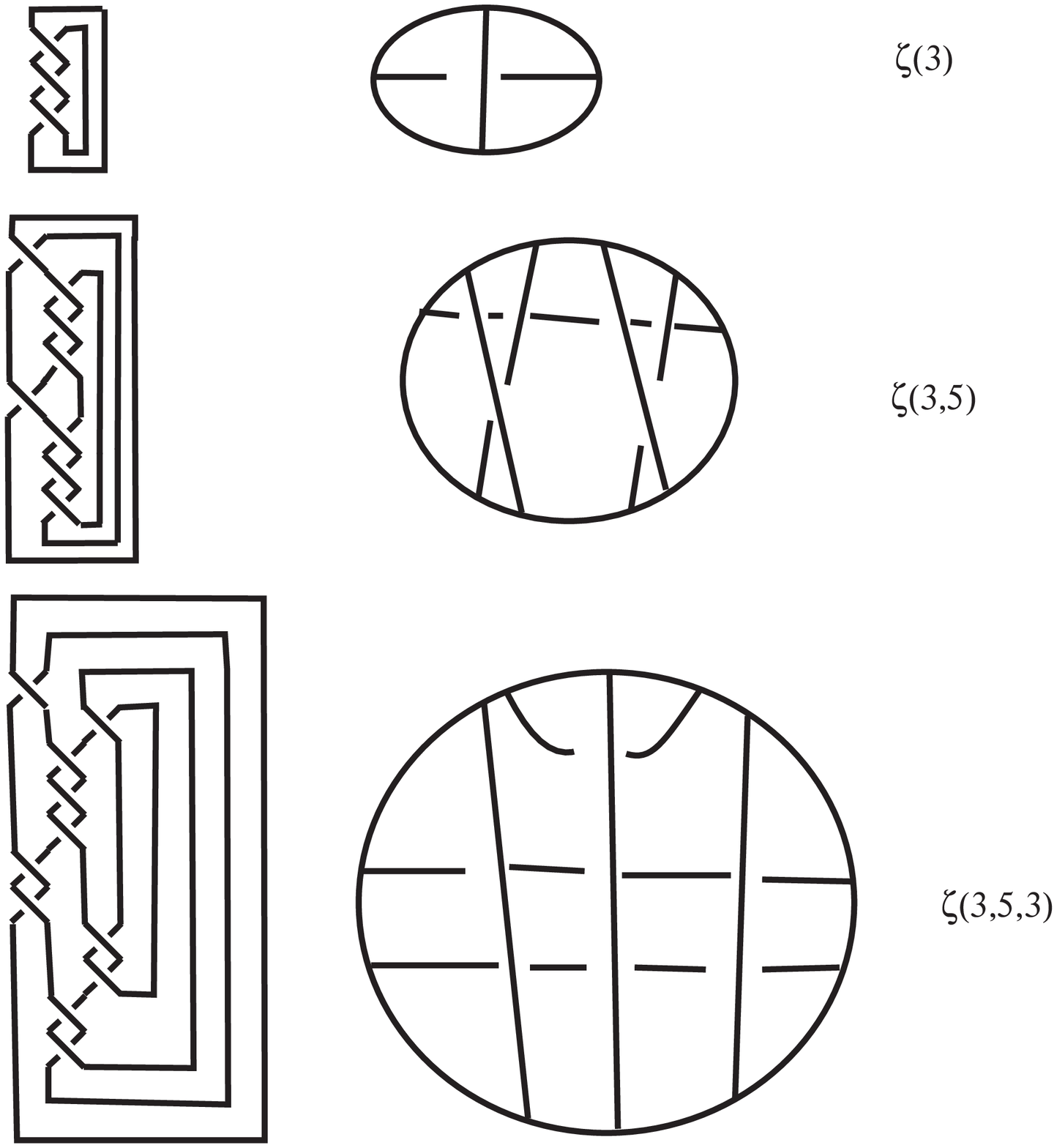}
\caption{Knots, Numbers, Diagrams. Some of the diagrams, knots and
  numbers which
are identified by now. While ladder diagrams give rational counterterms,
other topologies generate single, double and triple sums.}
\label{f2}
\end{figure}

This lead to a knot-to-number dictionary.
Fig.(\ref{f2}) summarizes some of the knot and number classes identified so far
\cite{pisa,bk15}.
Hence we can classify topologies by assigning knots to Feynman graphs
\cite{habil,pisa,plb}, and find that
graphs which deliver similar knots deliver similar transcendentals. This pattern works well
enough to abstract conjectures about the nature of these
transcendental numbers \cite{bk15,DaN}.
So far severe numerical tests have confirmed these
conjectures \cite{bbb}.


This allows to determine a search base of transcendental numbers
in which a diagram will evaluate. This knowledge suffices 
to turn a numerical answer
for a diagram into an analytic one, as was demonstrated 
for the first time in \cite{pisa}.
The rational weights with which the transcedentals seen in
counterterms
contribute depend on the specific nature of the theory realizing a
given
topology. It is a fascinating and still open problem to understand how
these weights can be derived from the representation theory of the Lorentz
group.


Meanwhile, we understand that the presence of extra symmetries like
gauge symmetries \cite{bdk} or supersymmetry \cite{john}
can annihilate the presence of certain transcendentals in the final result.


\section{The Hopf algebra of renormalization}
But then, if the above described primitive Feynman diagrams have such wonders in store,
what about general Feynman diagrams, with full-fledged subdivergences?
There are clearly patterns in the study of subdivergent graphs which are hardly
accessible with standard techniques of perturbation theory \cite{subdiv}.

\begin{figure}
\epsfysize=13cm\epsfbox{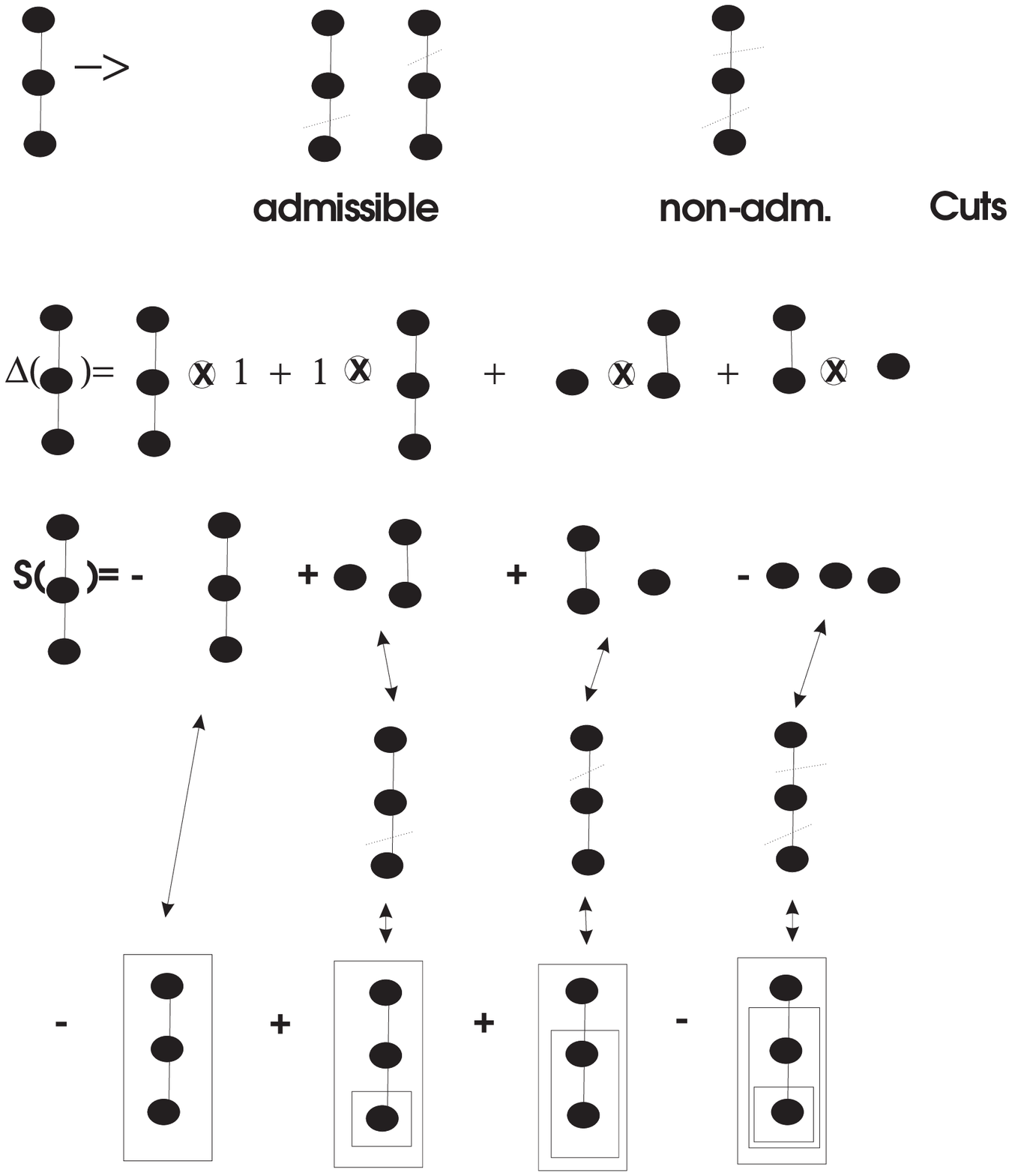}
\caption{The Hopf algebra of rooted trees. We define it using admissible
cuts
on the trees, and give the coproduct $\Delta$  in terms of admissible
cuts. An admissible cut allows for at most one single cut
in any path from any vertex to the root \cite{CK}.
Note that the antipode $S$ can  be formulated in terms of all cuts
\cite{CK},
and that cuts can be represented by boxes on the tree in the indicated
manner. The sign is determined by $(-1)^{n_c}$, where $n_c$ is the
number of boxes.}
\label{f3}
\end{figure}

It was the question how to disentangle a general Feynman graphs in terms of
primitive graphs which lead to the discovery of a Hopf algebra structure
to which we now turn \cite{hopf}. We change from the notion of
parenthesized words, used in \cite{hopf}, to the notion of
rooted trees. This notion turned out to be very convenient in the recent 
thorough investigation of the mathematical role
played by the Hopf algebra of renormalization \cite{CK}. 
Fig.(\ref{f3}) summarizes some basic
notions, as developped in \cite{hopf} and,
with particular emphasis given
to
rooted trees, in \cite{CK}.

\begin{figure}
\epsfysize=16cm\epsfbox{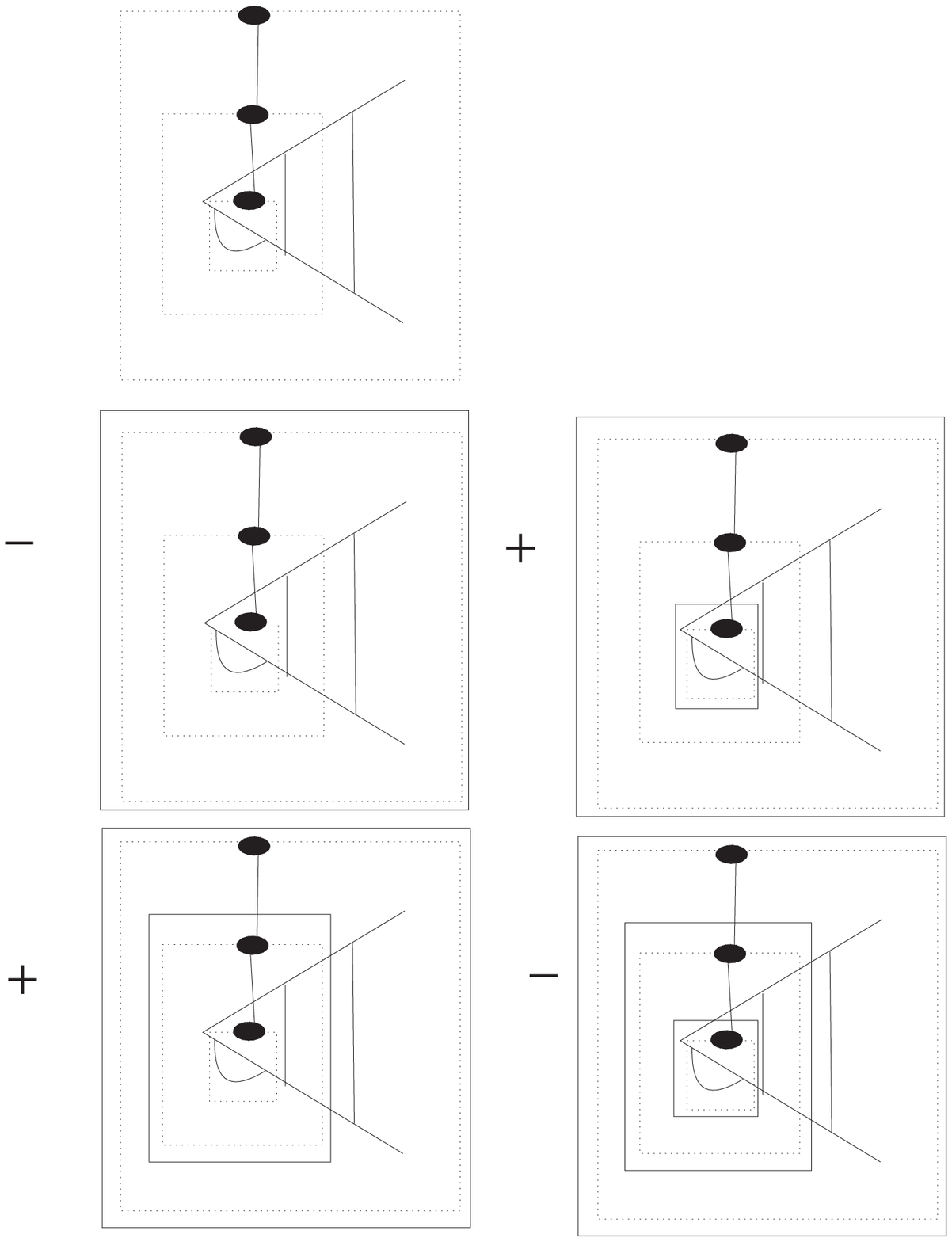}
\caption{The Hopf algebra of renormalization. We indicate how to
  assign a decorated tree
to a diagram. On such trees we establish the above Hopf algebra
structure.
Each black box corresponds to a cut on the tree, and these cuts
are in one to one correspondence with the forest structure.
We calculate the antipode on the tree, and represent the results
on Feynman diagrams, to find that the antipode corresponds to
the local counterterm.}\label{f4}
\end{figure}

Fig.(\ref{f4}) gives a diagrammatic explanation of this Hopf algebra, and shows how the
combinatorics of the forest formula derive from this Hopf algebra.
The figure explains that the local counterterm appears
as a fundamental  operation in the Hopf
algebra. It corresponds to the coinverse, the antipode. 
It reveals that the calculus of renormalization can be completely
understood
as derived from an underlying 
Hopf algebra structure. This includes the case of
overlapping
divergences, which can be resolved in terms of this Hopf algebra
using either Schwinger Dyson equations \cite{hopf},
powercounting arguments and differential equations for bare
Green functions \cite{CK} or algebraic methods \cite{new}.


The universal structure \cite{CK}
of the Hopf algebra of \cite{hopf} gives
hope that other diagrammatic expansions, for example 
asymptotic expansions, can be interpreted in terms of such algebraic
structures as well.


Let us also stress that the appearance of this Hopf algebra structure
in QFT establishes a link to recent developments in mathematics.
Quite the same Hopf algebra turned up in the work of Alain Connes
and Henri Moscovici \cite{CM}. The precise relation is now clarified
\cite{CK}. This gives a conceptual backing to the structure of local
quantum field theories which ought to be thoroughly investigated in
the future.
\clearpage
\section{Hopf algebra and four-term relations}
Let us close these remarks with a simple observation which
demonstrates the usefulness of the Hopf algebra in the understanding
of the appearance of transcendentals in Feynman diagrams.


The results of \cite{4t,bk4} point towards algebraic 
four-term relations between Feynman diagrams.
Clearly, to fully understand all relations between counterterms of diagrams, some
more work is needed. One might hope to construct a basis of diagrams which have
to be calculated, and then hopes to be able to determine all other diagrams
by knowing relations between diagrams.



\begin{figure}
\epsfysize=45mm\epsfbox{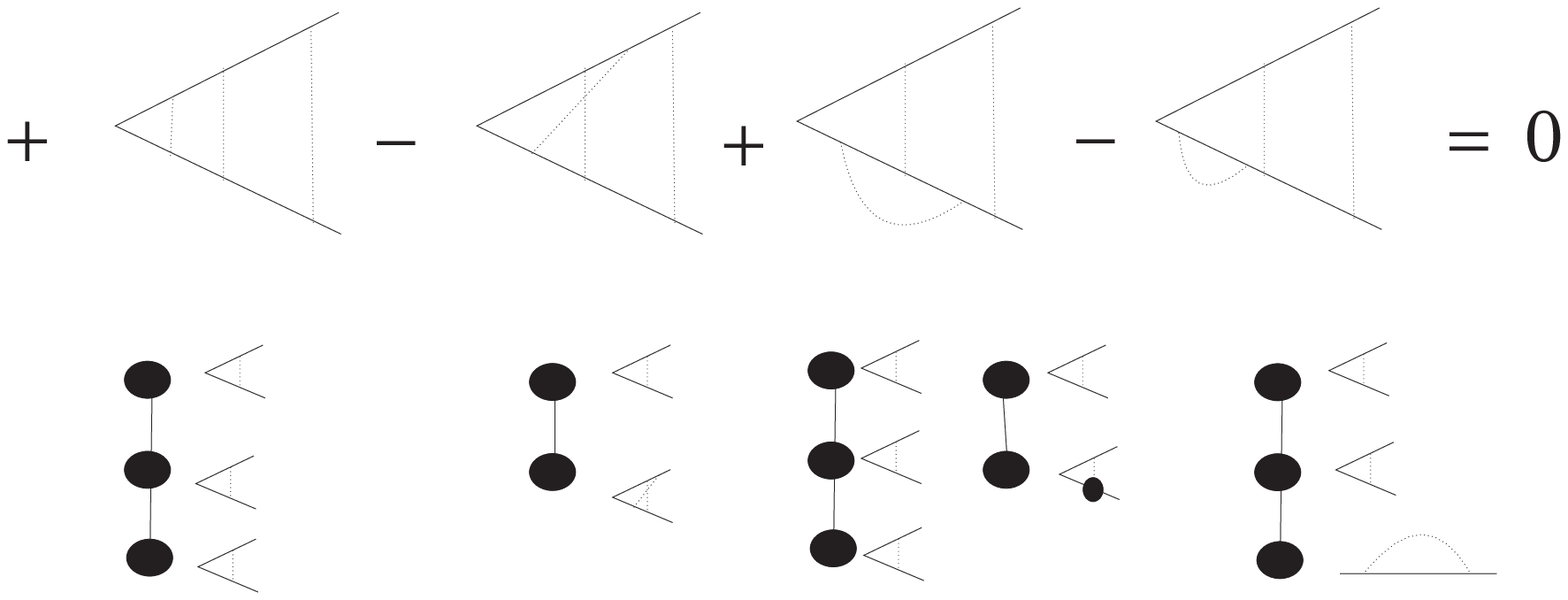}
\caption{Disentangling Feynman Diagrams in terms of primitive elements.
At the three loop level, a modified four-term relation
still has only two contributions which are of $\zeta(3)$ topology.
They are provided by the second and third graph above.
The third graph decomposes into two trees, when we decompose internal
vertex corrections into primitive elements.
Decomposing them into primitive elements is in accordance with a decomposition
into transcendental and rational parts. The transcendental parts
cancel
out as demanded by the four-term relation.}
\label{f5}
\end{figure}

The simplest example of a four-term relation
could appear at three loops, as at least two
chords are necessary, by definition.
It is given in Fig.(\ref{f5}). Relations between these four topologies are
obscured by the fact that there will be subdivergences present.

To overcome this problem we disentangle graphs
into primitive elements of the Hopf algebra of renormalization.
Let us give an idea how this might help to achieve
a full understanding of a (modified) four-term
relation
between diagrams.

We realize the four topologies in Yukawa theory.
Assume we expect that a four-term relation 
holds between the four graphs given in the figure, 
possibly modified
by terms including four-point couplings. 
Such a modified relation is suggested by a detailed study of the
results in \cite{4t}.
These two extra terms, as well as two
terms of the original four-term relation, are of ladder topology.


The two remaining terms have the $\zeta(3)$ topology, but a different structure as
rooted trees. Now let us decompose the one involving the one-loop
vertex correction as indicated in the figure. 
This corresponds to a decomposition
into primitive elements. 


Thus, for the four-term relation 
to hold, the only possible way is that the terms $\sim \zeta(3)$ cancel
on the level of antipodes.
This is indeed the case, as indicated in the figure, and will be reported in greater
detail elsewhere. Here, it suffices to note that a decomposition
into primitive elements of the Hopf algebra agrees with a decomposition
into the transcendental content of the various terms.
\section{Conclusions}
In this paper we succinctly reviewed 
a large body of results. Most of these results were gained
through an investigation of Feynman diagrams as entities of interest in their own
right.


Altogether, the underlying idea that the numbers seen in the calculation of Feynman diagrams
are not that arbitrary, but on the contrary determined by their topology,
underlies most of the results considered. 
The relation to braid-positive knots
comes as a surprise in the exploration of such ideas, and still awaits its final explanation.


The fact that there is a Hopf algebra structure underlying renormalization theory
points towards a beautiful connection between local quantum field theory and
the theory of operator algebras, which, in modern terminology, is
called non-commutative geometry.


If some of these ideas  come to fruition
in the future, 
we arrive at a picture of QFT which is quite different from what we find in textbooks
these days.


Let us start with an arbitrary Feynman graph. We first construct its decomposition
into primitive graphs, as determined by the Hopf algebra structure.
These primitive graphs come as letters in which our field theory is formulated.
We determine a minimal basis in these letters, guided by relations between graphs
as indicated by the study of a (modified) four-term relation.


The elements of this basis are Feynman graphs
which are connected to braid-positive knots, evaluating
to counterterms expressed in transcendental numbers, which,
in the best of all worlds, we could infer from the knowledge of a knot-to-number
dictionary gained from empirical data or some deeper insight in the future.
This dictionary must not stop at the level of counterterms.
The before mentioned results of David Broadhurst
on the finite part of diagrams \cite{DaI,DaH} indicate that the
story continues.



If we then solve one further problem, how to determine the rational numbers
which come as coefficients of these transcendentals, dependent 
on the spin representation
of the involved particles,  we approach an understanding of local QFT
which can be considered satisfactory.


Being in an optimistic mode, one might even wonder if a decomposition in terms
of primitives might have something to say about the asymptotic nature of the
perturbative expansion? The perturbation series is asymptotic at best,
but it might well be much better behaved if we disentangle it into rational
contributions, contributions $\sim\zeta(3)$, and so on, having in mind that each of these
numbers seems to refer to a different topology, which we might be forbidden to add up naively.


All of these properties seem to be restricted
to a local 'point-like' QFT. Indeed, it seems that most of the successes
attributed to theories
based on extended objects are present due to the fact that
these theories notoriously avoid the subtleties imposed on us
by the presence of quantum fields localized
at a point. These subtleties
are reflected by the presence of UV-divergent integrals. But then
 a careful study of
the
properties of these UV divergences indicate that they encapsulate an
enormously
rich structure at high loop orders, 
pointing towards the mathematics which might be needed
to extend our
understanding of QFT in the future.
\section*{Acknowledgements}
It is a pleasure to thank the organizers for a lively and stimulating
workshop. As always, I have to thank David Broadhurst for
an uncountable number of discussions 
and all his enthusiasm, and I thank Alain
Connes for guiding me into the realm of noncommutative geometry and
for
lot of illuminating discussions.
I thank the DFG for a support by a Heisenberg fellowship.

\end{document}